\documentclass[8.5pt,twoside,twocolumn]{article}

\usepackage[super,sort&compress,comma]{natbib} 
\usepackage{mhchem}
\usepackage{times,mathptmx}
\usepackage{sectsty}
\usepackage{balance} 
\usepackage{natbib}
\usepackage{hyperref}
\hypersetup{colorlinks=true,linkcolor=blue,citecolor=blue,urlcolor=blue}
\usepackage{graphicx} 
\usepackage{lastpage}
\usepackage[format=plain,justification=raggedright,singlelinecheck=false,font=small,labelfont=bf,labelsep=space]{caption} 
\usepackage{fancyhdr}
\usepackage{setspace}
\begin{document}

\thispagestyle{plain}
\fancypagestyle{plain}{
\renewcommand{\headrulewidth}{1pt}}
\renewcommand{\thefootnote}{\fnsymbol{footnote}}
\renewcommand\footnoterule{\vspace*{1pt}%
\hrule width 3.4in height 0.4pt \vspace*{5pt}} 
\setcounter{secnumdepth}{5}

\makeatletter 
\def\subsubsection{\@startsection{subsubsection}{3}{10pt}{-1.25ex plus -1ex minus -.1ex}{0ex plus 0ex}{\normalsize\bf}} 
\def\paragraph{\@startsection{paragraph}{4}{10pt}{-1.25ex plus -1ex minus -.1ex}{0ex plus 0ex}{\normalsize\textit}} 
\renewcommand\@biblabel[1]{#1}            
\renewcommand\@makefntext[1]%
{\noindent\makebox[0pt][r]{\@thefnmark\,}#1}
\makeatother 
\renewcommand{\figurename}{\small{Fig.}~}
\sectionfont{\large}
\subsectionfont{\normalsize} 

\fancyfoot{}
\fancyfoot[RO]{\footnotesize{\sffamily{1--\pageref{LastPage} ~\textbar  \hspace{2pt}\thepage}}}
\fancyfoot[LE]{\footnotesize{\sffamily{\thepage~\textbar\hspace{3.45cm} 1--\pageref{LastPage}}}}
\fancyhead{}
\renewcommand{\headrulewidth}{1pt} 
\renewcommand{\footrulewidth}{1pt}
\setlength{\arrayrulewidth}{1pt}
\setlength{\columnsep}{6.5mm}
\setlength\bibsep{1pt}


\twocolumn[
\begin{@twocolumnfalse}
\noindent\LARGE{\textbf{Detection of Gas Molecule using C$_3$N island Single Electron Transistor}}
\vspace{7pt}

\noindent\large{\textbf{S. Rani$^{1}$, S. J. Ray$^{1}$}}

\vspace{0.6cm}


\noindent \normalsize{C$_3$N is a recently discovered 2D layered material structurally similar to graphene, which has demonstrated immense prospect for future nanoelectronics. In this work, we have designed and investigated the operation and performance of a C$_3$N island single electron transistor (SET) for the first time. Using First-principles based calculations, we investigated the effect of various molecular adsorptions on the electronic and conduction behaviour of the SET. C$_3$N was found to be the perfect host material for capturing CO$_2$. The charge stability diagram carries the signature of different molecules within the SET and their presence can be uniquely identified from various line scans and normalised differential conductance behaviour obtained from it. Our results suggests the usefulness of such nanoelectronic structures for sensing toxic gas molecules which can be operational over a wide temperature range with detection sensitivity upto a single molecular level.}
\vspace{0.5cm}
\end{@twocolumnfalse}
]

\footnotetext{\textit{$^{1}$Department of Physics, Indian Institute of Technology Patna, Bihta 801106, India; E-mail: ray@iitp.ac.in, ray.sjr@gmail.com}}



\section{Introduction}
Two dimensional layered materials are crystalline nanomaterials with tuneable electronic, mechanical, optical and transport properties that are of interest for future nanoelectronics and spintronics. Graphene is the first discovered \cite{Novoselov2004} 2D material, but the absence of band gap reduces its application portability for switching. Doping it with Nitrogen and Boron could be an effective way to make it semiconducting which is desired in a Field-effect transistor (FET) structure. For this reason, Carbon(C)-Nitrogen(N) compounds with different N/C ratio is an area of huge interest due to their structural similarity to that of graphene. Out of the various combinations, monolayer polyaniline (C$_3$N) was found to be a hole-free, 2D layered structure with fascinating electronic \cite{Makaremi2017, Makaremi2018, Zhou2017, Zeng2018}, thermal \cite{Hong2018, Kumar2017}, mechanical \cite{Zhou2017} and chemical \cite{Li2017, Ma2018, Li2018, Preeti2018} properties, which was recently synthesised by direct pyrolysis\cite{Mahmood2016} and polymerisation \cite{Yang2017} process. In this structure, C$_3$N forms a hexagonal honeycomb lattice similar to that of Graphene with an indirect band structure, unlike Dirac Fermions \cite{Ray2017}. Recent reports indicated an ultra-high ON/OFF ratio in FET $\sim5.5\times10^{10}$ (significantly higher than any standalone 2D-layer system), carrier mobilities $\sim 180(220)$ cm$^2$V$^{-1}$s$^{-1}$ and existence of  spontaneous ferromagnetism at low temperature\cite{Yang2017} .

A single electron transistor (SET)\cite{Fulton1987} is a 3-terminal nanoelectronic structure comprised of Source (S), Gate (G) and Drain (D) electrodes. The channel region, as conventionally present in a FET, is replaced here with a quantum dot (QD) (or island). In the presence of tunnel barriers at the S/QD and D/QD junctions, the electronic conduction occours through the process of sequential quantum tunnelling while the gate is used to tune the electrostatic potential of the island\cite{RayPRA, RayWOLTE, Ray2016c} in a Coulomb blockade state. Scanning the conduction behaviour for various values of source-drain bias (V$_{d}$) and gate voltage ($V_{g}$) results in a rich phase diagram (known as charge stability diagram) that unveils the quantum transport behaviour of the system \cite{RayCPEM} which has potential applications in quantum metrology \cite{RayPRX}, sensing \cite{Ray2014a, Ray2014b, Ray2014c, Ray2015a, Ray2015b, Ray2016a, Ray2016b}, ON-chip circuit co-integration \cite{RayPRA} etc. In the recent times, 2D-materials were used for SET design due to the relative ease of fabrication and wider temperature range of operation, as demonstrated experimentally for graphene \cite{Stampfer2008, Barreiro2012, Connolly2010}. 

In the proximity of an adsorbed molecule, the resistance of a 2D material can change significantly \cite{Schedin2007}, owing to its high surface to volume ratio. It results in charge transfer and the sign of which determines the donor/acceptor nature of the adsorbent. In a confined nanostructure like a SET, this change can be higher due to quantum confinement and this property can be used for sensing as observed earlier \cite{Ray2016a, Ray2016b, Ray2018a}. In this work, we have used First-principle calculations to study the behaviour and and operation of a C$_3$N island SET device. We investigated the effect of the adsorption of various toxic gas molecules through parameters like charge transfer, charging and adsorption energy, density of states etc. on the characteristics of the SET and compared them with the behaviour observed for other 2D materials. By comparing the charge stability diagrams, line scans and conductivity behaviour for various island configurations, a powerful sensing procedure was proposed which can differentiate between the presence of individual gas molecules within the SET.

\begin{figure*}
\includegraphics[width=18cm]{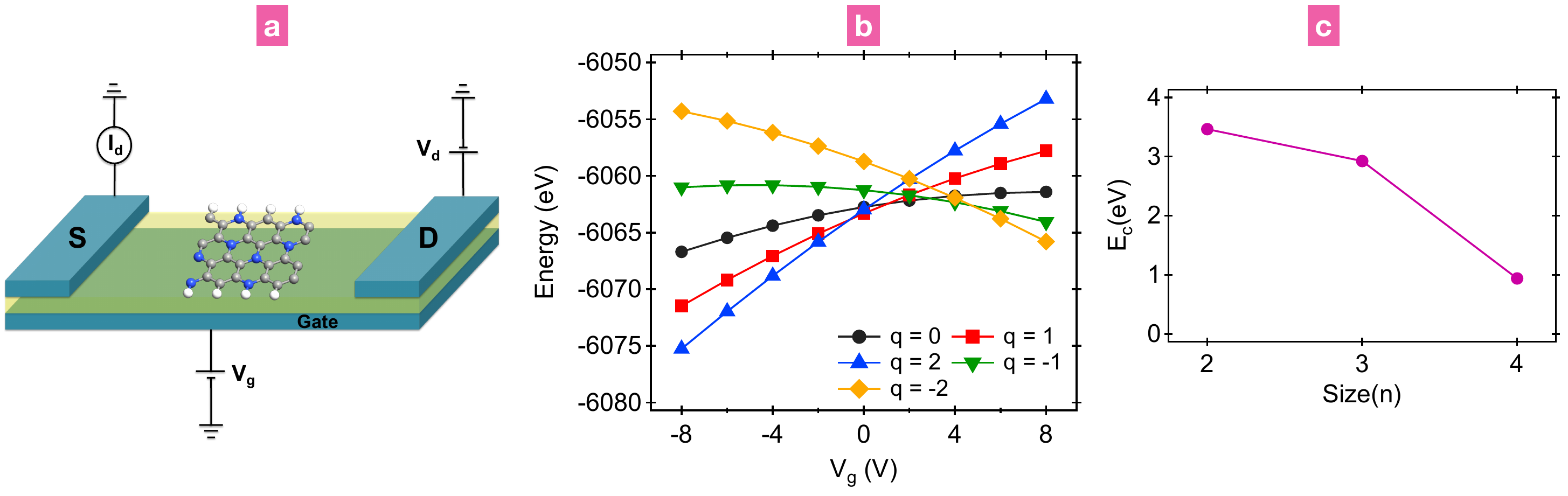}
\caption{{\small (a) Schematic of the C$_3$N island SET and (b) Gate voltage dependence of the total energy of the SET for different charge states (q) of the island, (c) Charging energy (E$_c$) as function of the size ($n$) of the C$_3$N island.}}
\label{fig.1}
\end{figure*}

\section{Computational details}
First-principles based Density Functional Theory (DFT) calculations are used to estimate the energies of the C$_3$N layer within the SET for different charge states. The Non-Equilibrium Green's Function (NEGF) based methodology used for this work was proposed here\cite{Brandbyge1999, Taylor2001, Brandbyge2002}, which for an equilibrium system inside a polarisable environment was used by Kaasbjerg and Flensberg \cite{Kaasbjerg2008}. Stokbro \cite{Stokbro2010} incorporated this with a DFT framework, as implemented within Atomistix Toolkit \cite{ATK}. Details of the associated electrostatics of the system can be found here\cite{Ray2015b}. Self-consistent calculations are performed under the generalised gradient approximation (GGA) of the Perdew-Burke-Ernzerhof (PBE) \cite{PBE} exchange correlation functional, where the wave functions are expanded within a double-$\zeta$ polarised (DZP) basis set. To account for the Van Der Waal's interaction, semi-empirical corrections by Grimme D2 \cite{Grimme2006} and D3 \cite{Grimme2010} were included within the GGA. For solving the Poisson's equation, metallic electrodes are considered to be equipotential surfaces subject to Neumann's boundary conditions.  For estimating the density of states (DOS), a Monkhorst-pack\cite{Monkhorst} $k$-point grid of 8$\times$8$\times$1 was used and spin-polarisation was considered within the exchange correlation functional to include the magnetic contribution of different molecules, whenever required. To establish the most equilibrium configuration, all the structures were allowed to relax fully until the force on each atom is smaller than 10$^{-3}$ eV/{\AA}.

\section{System Description}
Schematic of the SET device used for the present work with C$_3$N island is illustrated  in Fig.~\ref{fig.1}(a). I$_{d}$ represents the source-drain current at a given V$_{d}$ and V$_g$. The island of the SET was made of a $4\times4$ supercell of C$_3$N monolayer, edge-passivated with H-atoms. The outer box size of the SET was 14{\AA} $\times$ 14{\AA} $\times$ 21{\AA} which included the source, drain and gate electrodes with the following dimensions: 14{\AA} $\times$ 8.3{\AA} $\times$ 4{\AA}, 14{\AA} $\times$ 8.3{\AA} $\times$ 4{\AA}, 14{\AA} $\times$ 1{\AA} $\times$ 21{\AA} respectively. C$_3$N has a planar structure which was kept parallel to the bottom gate backed by a dielectric layer of thickness 4.7\AA\, and dielectric constant 10$\varepsilon_{o}$ ($\varepsilon_{o}$ is the permittivity of free space). Gold was considered to be the electrode material for S, D and Gate while transport direction was considered along the zigzag axis of C$_3$N.

\section{Results and Discussion}

\subsection{Electrostatics}

\begin{figure*}
\includegraphics[width=18cm]{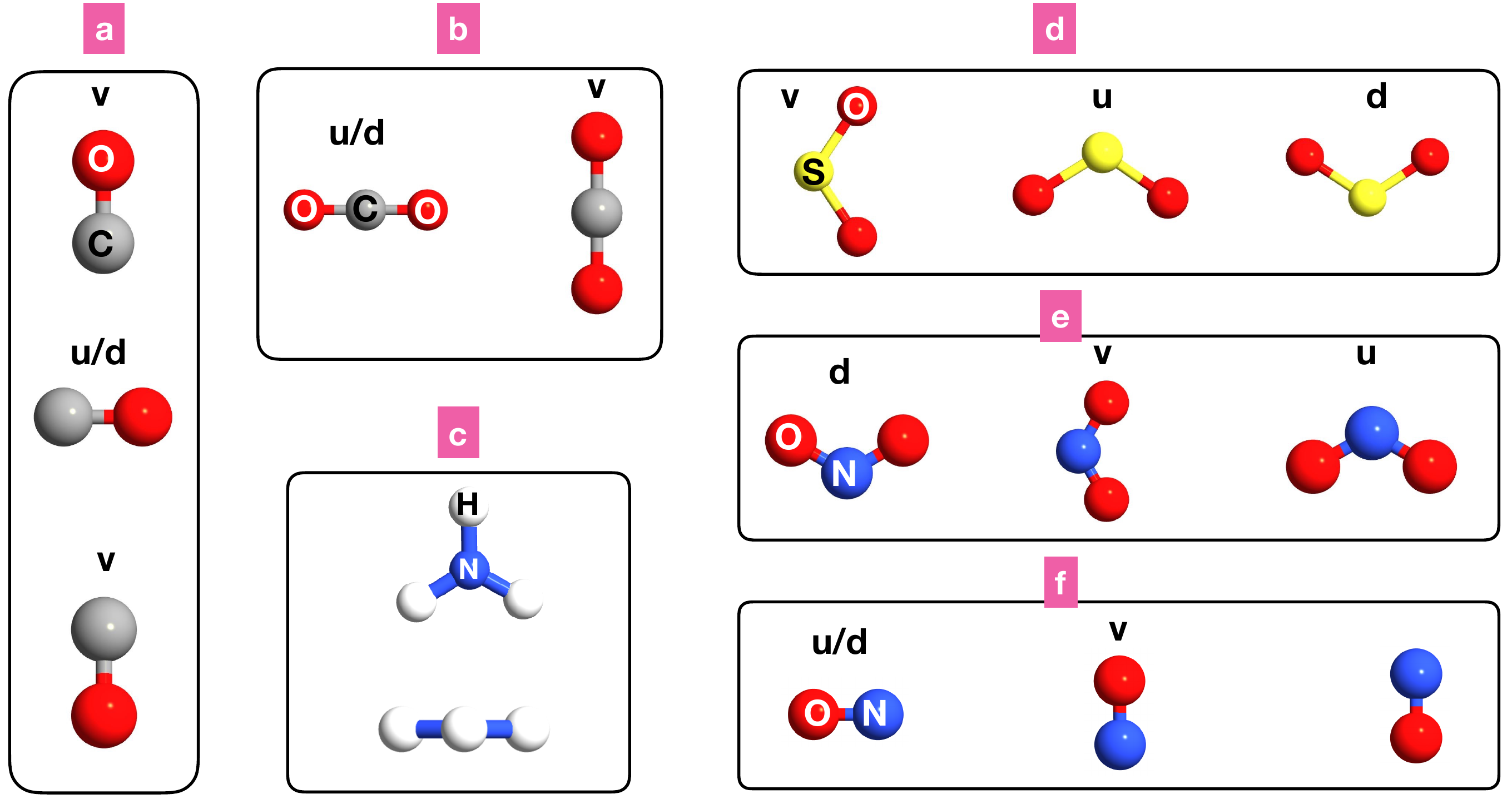}
\caption{{\small Schematic of the orientations of different gas molecules with respect to the C$_3$N layer  within the SET : (a) CO, (b) CO$_2$, (c) NH$_3$, (d) SO$_2$, (e) NO$_2$ and (f) NO.}}
\label{fig.2}
\end{figure*}

The total energy ($E$) of the C$_3$N island within the SET as function of $V_{g}$ is illustrated in Fig.~\ref{fig.1}(b). For different charge states $q$, $E$ can be analytically described\cite{Ray2014a} using Eqn.~\ref{eqn.1}
\begin{equation}
E (q, V_g) = E_{o} + \alpha qV_{g} + \beta (eV_{g})^{2} + qW
\label{eqn.1}
\end{equation}
Here $E_0$ is the constant term in energy and $W$ is the work function of the metallic gold electrode with a value of 5.28 eV. $\alpha$ and $\beta$ represent the numerical coefficients that are estimated through least-square fitting of Eqn.~\ref{eqn.1} to the DFT calculated energies. The solid lines in Fig.~\ref{fig.1}(b) represent the fitted values of energy from which the values of the coefficients were estimated to be : $\alpha$ =  0.56 and $\beta$ = 0.0045 $eV^{-1}$. The electrostatic coupling between the gate and the island is quantified through $\alpha$ which depends on the relative position/orientation between the island and the gate. The value of $\beta$ on the other hand is an estimation of the electrostatic polarisation contribution of the island towards the total energy, which is significantly smaller in magnitude than the linear term. This indicates that electrostatic operation of the SET is primarily gate dominated compared to the electrostatic polarisation contributions, which is expected for a planar island structure as used in the present case.  

Further insight in the conduction behaviour of the SET \cite{Stokbro2010, Ray2014a, Ray2015b} can be obtained from the charge stability diagram as shown in Fig.~\ref{fig.5}(a). Subject to the satisfaction of the inequality in Eqn.~\ref{eqn.2},
\begin{equation}
e|V_{d}|/2 > (E_{c} + W) > -e|V_{d}|/2
\label{eqn.2}
\end{equation}
diamond shaped regions are observed on the $V_{d} - V_{g}$ plane, where $E_{c}$ is the charging energy. The height of the central diamond corresponds to the charging energy of the island in its ground state, which for C$_{3}$N is found to be 1.68 eV. The dependance of $E_{c}$ for different size ($n\times n$) of the monolayer C$_{3}$N island is shown in Fig.~\ref{fig.1}(c). For a fixed S/D separation, the increase in the size of the island ($n$) indicates an enhancement in the capacitive coupling between the S-island-D and a reduction in E$_c$, which arises due to quantum confinement. Near V$_d$ = 0 and V$_g$ = 0 in Fig.~\ref{fig.5}(a), non-zero conduction occours around V$_d$ = $1.02$V suggesting the ON/OFF operation in this SET is possible through a small voltage swing.

Due to the quantised nature of conduction, the charge state of the island gets significantly influenced in the proximity of an external molecule, which gets reflected in the charge stability diagram. For the present case, the effect was investigated for six different molecules namely : (a) CO, (b) CO$_{2}$, (c) NH$_{3}$, (d) NO, (e) NO$_{2}$ and (f) SO$_{2}$ that are commonly found in the atmosphere. In order to determine the most favourable  configuration of molecular adsorption on the monolayer, several different orientations of each molecule are considered namely (i) up ($u$), (ii) down ($d$) and (iii) vertical ($v$) as illustrated in Fig.~\ref{fig.2}. The strength of the binding can be estimated from the adsorption energy ($E_{a}$) defined by Eqn.~\ref{eqn.3},
\begin{equation}
E_{a} = E_{layer} + E_{molecule} - E_{layer+molecule}
\label{eqn.3}
\end{equation}
where $E_{layer}$ and $E_{molecule}$ are the energies of the 2D layer and the molecule respectively while $E_{layer+molecule}$ is the energy of the combined layer and molecule system in its equilibrium configuration within the SET. 

\begin{figure*}
\includegraphics[width=14cm]{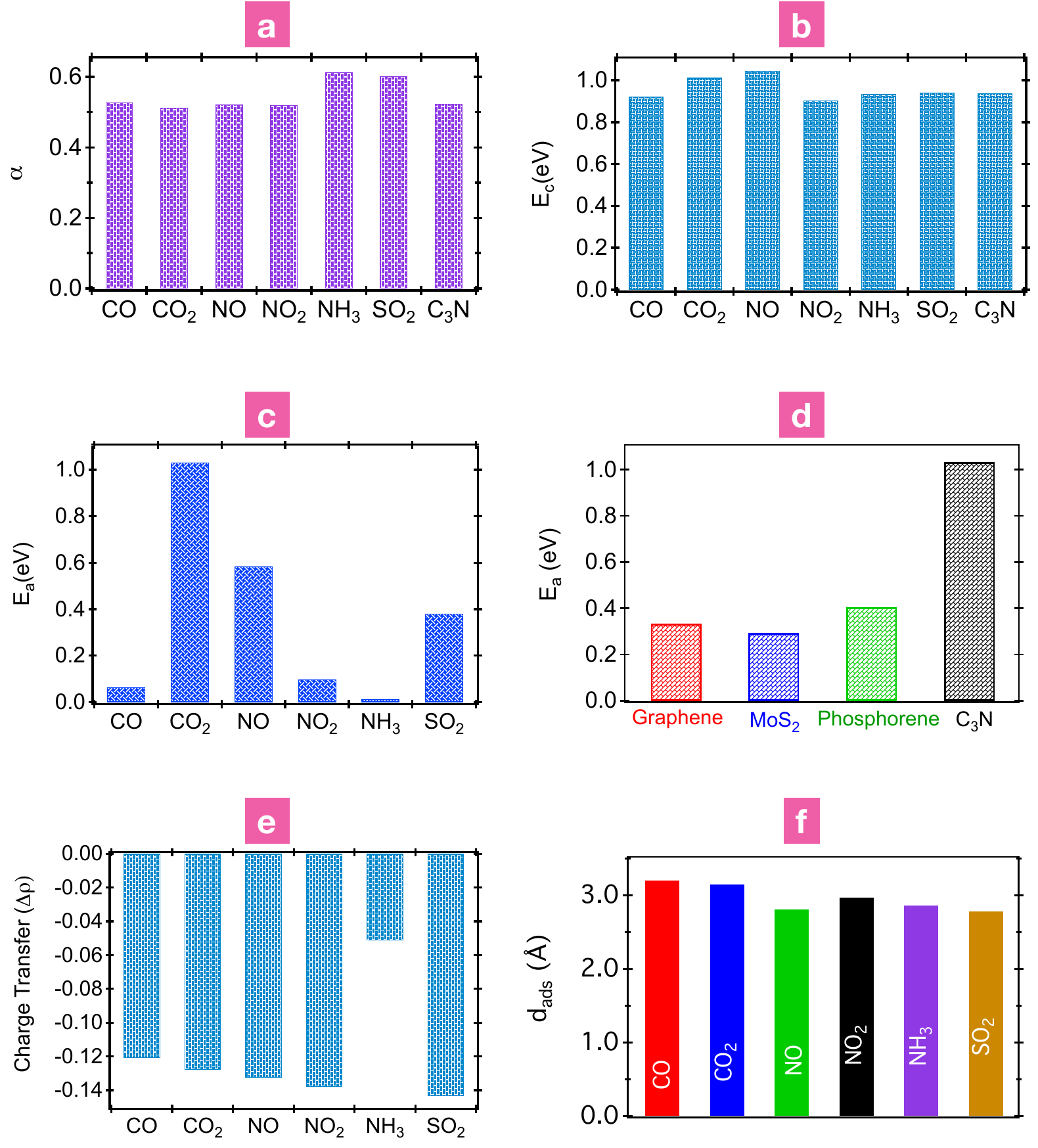}
\caption{{\small Variation of (a) $\alpha$, (b) E$_c$, (c) E$_a$, (e) $\Delta\rho$, (f) d$_{ads}$ for different gas molecules adsorbed on the C$_3$N layer within the SET,  (d) Change in E$_a$ for CO$_2$ adsorbed on different 2D layers in similar SET configurations.}}
\label{fig.3}
\end{figure*}

In the presence of the molecule, the nature of the electrostatics within the SET can be quantified through the estimation of $\alpha$ as illustrated in Fig.~\ref{fig.3}(a). Overall, the value of $\alpha$ stayed within the range of 0.52 - 0.63 with the maximum for NH$_{3}$  adsorbed configuration. In all cases, $\alpha$ is significantly higher than the corresponding $\beta$, therefore concluding that the gate dominance stays prominent in the all these configurations to control the electrostatic operation. The charging energy $E_c$ on the other hand describes the change in the conduction behaviour in different island configurations as shown in Fig.~\ref{fig.3}(b). Similar to $\alpha$, $E_c$ does not go through major variations in different island configurations. The maximum was observed for NO adsorbed configuration, which is about 0.1 eV higher than the pristine configuration and the values of $E_c$ lies within the range of 0.9-1eV suggesting that a typical $V_d\sim$ 1V is sufficient in the current SET for getting a finite conduction from the OFF state. Combining the behaviour of $\alpha$ and $E_c$ in various SET, it can be said that the nature of electrostatics and conduction do not go through major changes in the presence of a molecule close to the island.

The adsorption energy $E_{a}$ estimated for different gas molecules in their most equilibrium configurations on the C$_3$N are summarised in Fig.~\ref{fig.3}(c). Significant dispersions are observed in the $E_{a}$ behaviour with the highest for CO$_2$ at a value of 1.03 eV. To correct for the basis set superposition error (BSSE), counterpoise correction\cite{Boys1970} was used on the LCAO\footnote{Linear combination of atomic orbitals} basis set resulting in an $E_{a}$ of 0.96 eV for CO$_2$ which matches very well with the earlier value. The CO$_2$ molecule prefers to align itself in the $u/d$ orientation (Fig.~\ref{fig.2}b) in its most stable adsorption configuration. CO$_2$ molecule has a large electric quadrupole moment originating from the strong dipolar C=O bond. Within the electrostatic environment of the SET, it offers large polarisability and strong interaction between the molecule and the C$_3$N layer resulting in large $E_{a}$. It is significantly higher than the $E_a$ values estimated for Graphene (194\%), MoS$_2$ (240\%) and Phosphorene (157.5\%) in similar SET environment as shown in Fig.~\ref{fig.3}(d). Schematics of the preferred adsorption locations of various gas molecules on the C$_3$N can be found in Fig. S1, S2, while for other 2D layers in Table S1 in the Supporting Information section\cite{SI-C3N-SET}. For NO and SO$_2$, the $E_a$ values stay within range of 0.38 - 0.6 eV.  It was observed for SO$_2$ that in the adsorbed configuration, primary areas of electrons concentration occours around the C$_3$N layer with depletion from the O-atoms which explains the significant $E_{a}$ observed here. While looking at the molecular orientation, $u/d$ is the preferred alignment for NO and $d$ was found to be the equilibrium orientation for SO$_2$ (Fig.~\ref{fig.2}d, f). The values of $E_{a}$ are significantly small for CO, NO$_2$ and the lowest value was observed for NH$_3$. Comparing these values with other 2D materials in a similar electrostatic environment, it was observed that $E_{a}$ for CO is almost 98.4\% less than graphene and more than 4 times lower than MoS$_2$ and Phosphorene. Similarly for NO$_2$ and NH$_3$, E$_a$ is significantly lower than that of  graphene. Among all the molecules considered in the present investigation, the adsorption strength for CO$_2$ is very high suggesting that C$_3$N can be a good candidate for detecting CO$_2$ which also agrees with the high CO$_2$ uptake capacity reported through Grand Canonical Monte Carlo Calculations \cite{Li2017}. The adsorption strength for NO and SO$_2$ are also large, suggesting the usefulness of C$_3$N for sensing these molecules. 

To determine the interaction between the gas molecules and the host material, charge transfer plays a crucial role which at a location $\vec{r}$ can be quantified using Eqn.~\ref{eqn.4}, 
\begin{equation}
\Delta\rho(\vec{r}) = \rho_{host+mol}(\vec{r}) - \rho_{host}(\vec{r}) - \rho_{mol}(\vec{r})
\label{eqn.4}
\end{equation}
where $\rho_{host+mol}(\vec{r}), \rho_{host}(\vec{r})$ and $ \rho_{mol}(\vec{r})$ represent the respective charges of the adsorbed (host layer+molecule) assembly, 2D layer and the molecule in their most equilibrium configurations. We observed a donor type contribution from all the gas molecules studied here as illustrated in Fig.~\ref{fig.3}(e). The highest value of $\Delta\rho\sim$ -0.142$e$ was observed for SO$_2$ while $\Delta\rho$ values stay above -0.12$e$ for all other molecules except NH$_3$. This indicates that in the presence of these molecules, significant changes in the resistivity should be observed in an experiment. The distribution of charges in the C$_3$N layer and various gas molecules after adsorption can be found in Fig. S4 and Highest occupied molecular orbital (HOMO) of the gas molecules are shown in Fig S3. The large value of $\Delta\rho$ for CO$_2$ agrees well with the trend observed in $E_a$ for these molecules. The equilibrium distance of separation ($d_{ads}$) for different molecules on C$_3$N is shown in Fig.~\ref{fig.3}(f). The maximum of $d_{ads}$ was observed for CO ($\sim$ 3.2\AA) and the minimum for SO$_2$ ($\sim$ 2.78\AA). For NO and SO$_2$, the values are comparable and combining them with larger values of $E_a$, suggests that C$_3$N could be a good host material for adsorbing these two types of gas molecules. For all molecules, there was no evidence of any physical bond formation between the molecule and the C$_3$N layer, suggesting the physio-adsorption nature of adsorption process involved in these cases and no major modifications of the the surface structures of the host layers were observed. On the C$_3$N layer, $d_{ads}$ is 17.36\% (for NO$_2$), 8.25\% (for CO$_2$) higher than MoS$_2$ while for CO, it is 12.2\% higher compared to Phosphorene. Thus considering the high value of $E_a$ and lower value of $d_{ads}$, it can be concluded that Phosphorene is a preferred material for adsorbing NO$_2$ and CO compared to C$_3$N.

\begin{figure}
\includegraphics[width=9cm]{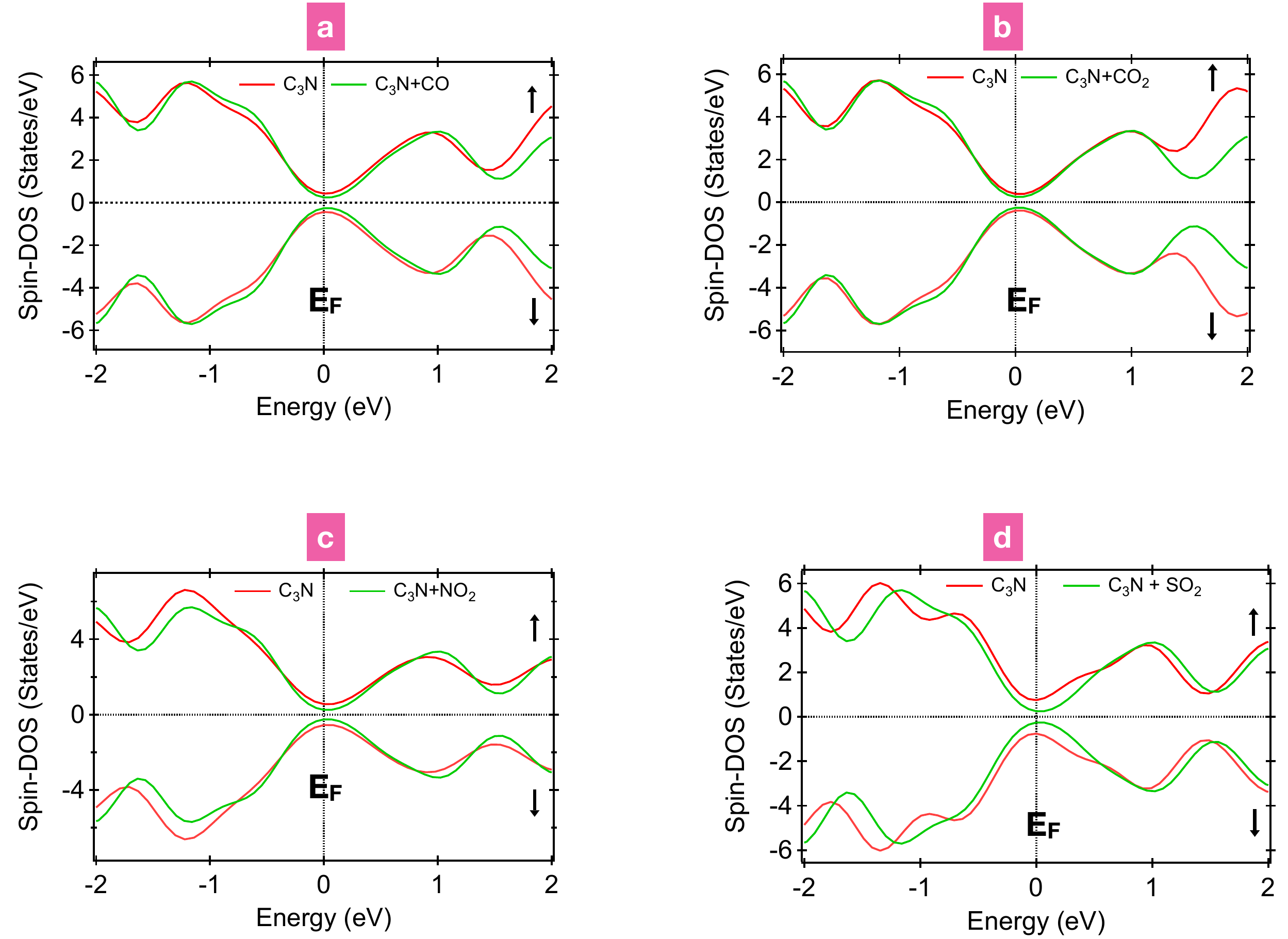}
\caption{{\small Spin density of states of the C$_3$N layer in pristine and adsorbed configuration with (a) CO, (b) CO$_2$, (c) NO$_2$ and (d) SO$_2$ molecules. E$_F$ is the Fermi Energy.}}
\label{fig.4}
\end{figure}

\begin{figure*}
\includegraphics[width=16cm]{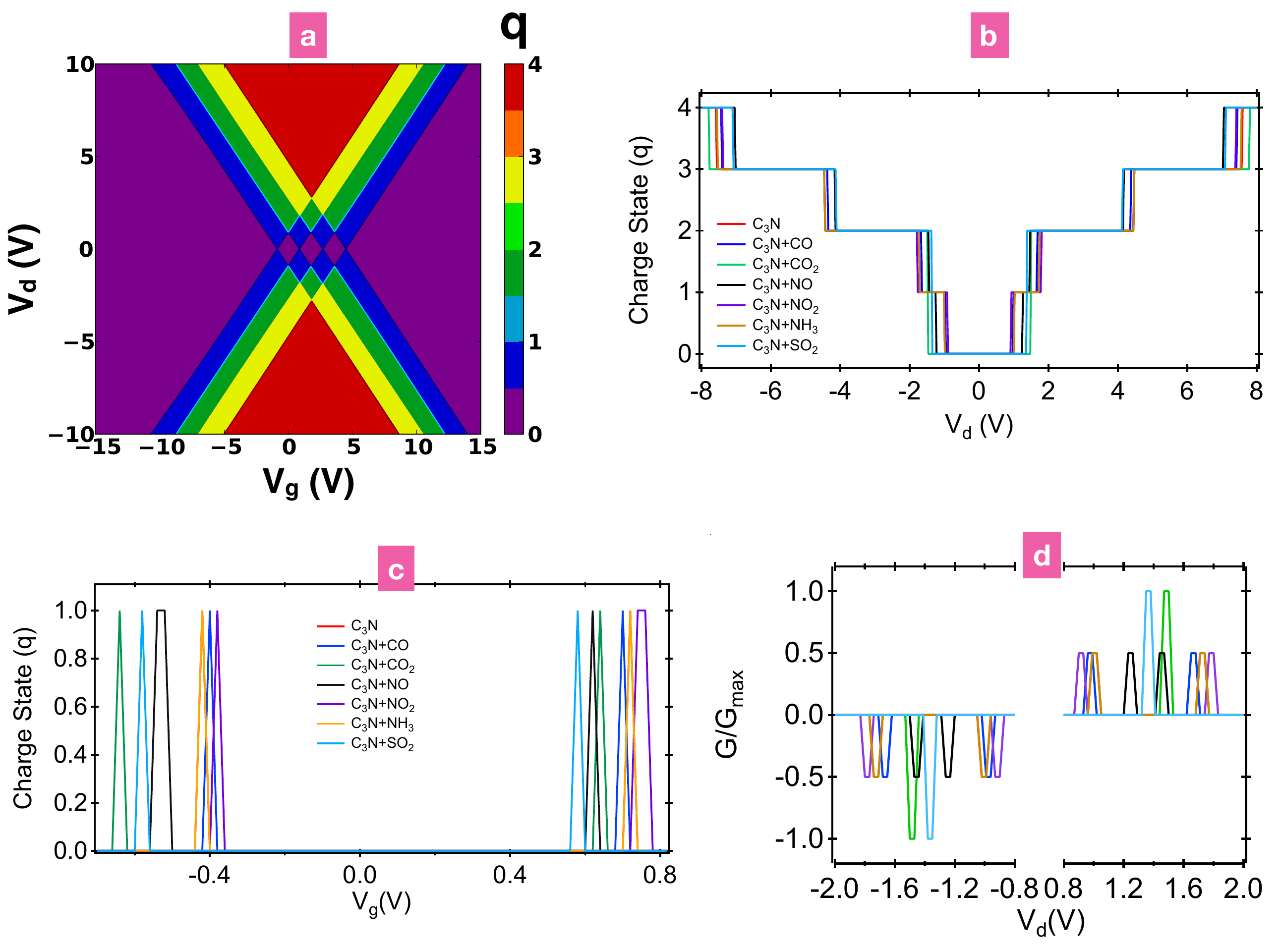}
\caption{{\small (a) Charge Stability Diagram of a C$_3$N based SET (colorbar represents different charge states $q$) and line scans taken along (b) V$_g$ and (c) V$_d$ for various island configurations with different gas molecules, (d) Normalised differential conductance corresponding to (c) as function of V$_d$ for these adsorbed configurations.}}
\label{fig.5}
\end{figure*}

Except NH$_3$, $E_{a}$ values of all the other molecules investigated here are found to be significantly large compared to the thermal background available at room temperature. At a given temperature T,  the desorption rate of a gas molecule is proportional to $\sim exp(-E_{a}/k_BT)$. For CO$_2$, NO and SO$_2$, the minimum temperature required to overcome the adsorption barrier is $\sim$ 1000K. This suggests that for these molecules, C$_3$N can be safely used over a large temperature range.


To understand the effect of gas molecule adsorption on the electronic properties of C$_3$N, the density of states (DOS) were calculated in the pristine and the adsorbed cases as shown in Fig.~\ref{fig.4}. In the presence of electric field, the semiconducting nature of the host layer gets modified which can be expected for a 2D semiconductor with a small semiconducting gap due to the changes in conduction and valance band positions \cite{Liu2015, Kim2015}. The DOS profile here stays symmetric around the Fermi level ($E_{F}$) irrespective of the adsorption conditions. In the presence of the gas molecule, the DOS behaviour do not go through any significant changes around the $E_{F}$. However, minor modifications can be observed at higher energies on both sides of $E_{F}$. For CO and CO$_2$ adsorptions some additional states can be found at around 2 eV while for NO$_2$ and SO$_2$, similar features can be seen between -1 eV to -2 eV. In the spin specific DOS profiles, no significant differences are observed between the spin-$\uparrow$ and spin-$\downarrow$ contributions, suggesting the absence of any paramagnetic influences from the molecules in this specific scenario. Overall, the shape of the DOS profile stays uninterrupted due to the molecular adsorption and presence of additional states at higher energies can be attributed to the charge transfers between the molecule and the host layer. 
 
We have also investigated the effect of multiple adsorption by considering two situations : (a) multiple molecules for a fixed sized host layer and (b) single molecule on a host layer of varied sizes. In both these cases, no major modifications in the overall DOS behaviour was observed except the appearance of few additional energy peaks like single molecule adsorption case, suggesting that the electronic properties of C$_3$N do not get affected significantly as a result of the adsorption processes.

\subsection{Detection Methodology}

The charge stability diagram (CSD) of the SET with an island structure comprised of C$_3$N layer adsorbed with SO$_2$ molecule is illustrated in Fig.~\ref{fig.5}(a). For different values of the V$_d$ and V$_g$, detailed information about the charge state of the island can be obtained from it and such a diagram is sensitive to the presence of a particular molecule within the SET. Therefore it can be used to uniquely identify the presence of a particular molecule adsorbed onto the host layer. This can be made more clear by taking the line scans from such CSD at different values of V$_d$ and V$_g$ as shown in Fig.~\ref{fig.5}(b-c). In the line scans taken along V$_d$ (for a fixed V$_g$) in Fig.~\ref{fig.5}(b), step like features can be observed for various island configurations symmetrically placed around V$_d$ = 0V. The symmetric nature indicates the symmetry of the tunnel barriers at S/island and D/island junctions. Starting from V$_d$ = 0V, with an increase in the V$_d$ the charge state of the island moves to its next excited state and subsequently the conduction stays blocked for that region of V$_d$ due to the Coulomb Blockade effect. The charging energy in various excited states can be estimated from the width of such plateaus. 

For a fixed V$_d$, line scans taken along V$_g$ for different island configurations are shown in Fig.~\ref{fig.5}(c). At a small and fixed $V_{d}$, line scans along $V_{g}$ resulted in the occurrence of periodic charge peaks while individual charges gets added/removed from the island. For various molecules such charge peaks of equal heights arise at various values of V$_g$ which can be distinguished by their respective coloured lines. Similar differences in the line scans are also visible in Fig.~\ref{fig.5}(b), where the V$_d$ values at which various plateaus arise are different for different molecules within the SET. This feature can be used to distinguish between different gas molecules in a C$_3$N island SET by comparing between respective line scans taken along various V$_d$/V$_g$ axes. 

In an experimental situation, the charge stability diagram is commonly obtained from the current (or conductance) measurements \cite{RayPRX, RayCPEM} on the 2D voltage plane. The charge peaks are replaced by current (or conductance) peaks and the general trends/shapes will vary accordingly. For comparison, the normalised differential conductance behaviour estimated from the representative line scans in Fig.~\ref{fig.5}(b) are illustrated in Fig.~\ref{fig.5}(d). Corresponding to different island configurations, the conductance peaks arise at different values of V$_g$ and peaks of a particular order are clearly distinguishable for different molecules. The uniqueness of the charge stability diagrams for specific island structure assures that such conductance peaks will be sensitive to the presence of a specific molecule, which can be used for sensing such gas molecules using a C$_3$N island SET. The first order conductance peaks appear below V$_d$ = 2V which suggests the low power operation in this case. The conductance peaks for SO$_2$ and CO$_2$ molecules are larger in heights compared to the others indicating that an enhanced resolution is available for their detection. At these particular values of V$_d$, the change in the charge state of the island for SO$_2$ and CO$_2$ is double than the others, which is the reason behind this. In the negative V$_d$ region of Fig.~\ref{fig.5}(d), the change in current $>0$ while $\Delta V_d<0$, which explains the origin of negative conductance peaks. The device dimensions used in this work can be scaled for experimental works and are expected to offer similar operational behaviour. In practice, alumina could work as a dielectric material of choice due to its moderate dielectric constant ($\varepsilon \sim 9-11\varepsilon_o$) as also used for similar systems. In order to increase the tunnelling current, an array of such SETs could be used as demonstrated earlier \cite{Ray2015a}.

\section{Conclusion}

In this work, we have used the NEGF combined DFT methodologies to understand the operational behaviour of a newly discovered 2D material, C$_3$N island SET device. The electrostatics of such SET in the pristine and adsorbed configuration of the island is controlled primarily by the gate. The adsorption energy was found to be significantly large for CO$_2$, NO and SO$_2$ and maximum for CO$_2$, indicating C$_3$N to be a superior host material compared to Graphene, Phosphorene and MoS$_2$ for CO$_2$. The small charging energy of the SET in its ground state offers a low power operation of this device between the ON/OFF states. The electronic and structural properties of the host layer stays almost unaffected as a result of adsorption and no evidence of paramagnetic contribution was observed in the spin density of states as a result of the Physio-adsorption of molecules. The presence of different molecules within the SET can be identified from the charge stability diagram and various line scans taken at specific values of V$_d$ and V$_g$.  The normalised differential conductance estimated here clearly distinguishes between the different molecules with distinct peaks appearing at specific values of V$_d$. This can be used for sensing such inorganic molecules, which for the present case was found be to be additionally sensitive to the presence of CO$_2$ and SO$_2$. Due to the enhancement of air pollution level globally, presence of toxic gas molecules in atmosphere is a matter of huge concern. We believe that our current findings will provide a new route for the fast/rapid detection of such molecules in an energy efficient architecture. 

\section*{Acknowledgments}

This work was financially supported by Department of Science and Technology, Govt. of India through the  INSPIRE scheme (Grant Ref: DST/INSPIRE/04/2015/003087).

\footnotesize{
\bibliography{Bibliography-ATK} 
\bibliographystyle{rsc} 
}

\end{document}